\documentclass[sigconf]{acmart}

\AtBeginDocument{%
  \providecommand\BibTeX{{%
    \normalfont B\kern-0.5em{\scshape i\kern-0.25em b}\kern-0.8em\TeX}}}

\acmSubmissionID{123-A56-BU3}

\usepackage{amsmath,amsfonts}
\usepackage[ruled]{algorithm2e}
\usepackage{graphicx}
\usepackage{textcomp}
\usepackage{xcolor}

\usepackage{lineno,hyperref}

\modulolinenumbers[5]

\usepackage{mwe}  
\usepackage{grffile}
\usepackage{hyperref}
\usepackage{url}
\usepackage{graphicx}
\DeclareGraphicsExtensions{.eps,.pdf}
\usepackage{multirow}
\usepackage[ruled]{algorithm2e}
\usepackage{amsmath}
\usepackage{pifont}
\usepackage{enumerate}
\usepackage{tikz}
\usetikzlibrary{calc,backgrounds,arrows,trees,shapes,snakes,shadows,patterns,matrix,fit,patterns,positioning,automata,decorations.pathmorphing,decorations.pathreplacing}
\usepackage{epstopdf}
\usepackage[aboveskip=2pt,belowskip=-2pt]{subcaption}
\usepackage[font=footnotesize]{caption} 
\usepackage[ruled]{algorithm2e}
\SetKwRepeat{Do}{do}{while}

\newcommand{\mathtext}[1]{\mathrm{\textit{#1}}}

\copyrightyear{2020}
\acmYear{2020}
\setcopyright{acmcopyright}\acmConference[ICPP '20]{49th International Conference on Parallel Processing - ICPP}{August 17--20, 2020}{Edmonton, AB, Canada}
\acmBooktitle{49th International Conference on Parallel Processing - ICPP (ICPP '20), August 17--20, 2020, Edmonton, AB, Canada}
\acmPrice{15.00}
\acmDOI{10.1145/3404397.3404441}
\acmISBN{978-1-4503-8816-0/20/08}

\begin{document}

%
\title{Enabling performance portability of data-parallel OpenMP applications on asymmetric multicore processors}

\author{Juan Carlos Saez}
\affiliation{\department{Facultad de Inform\'atica}
\institution{Complutense University of Madrid}
}
\email{jcsaezal@ucm.es}

\author{Fernando Castro}
\affiliation{\department{Facultad de Ciencias F\'isicas}
\institution{Complutense University of Madrid}
}
\email{fcastror@ucm.es}

\author{Manuel Prieto-Matias}
\affiliation{\department[0]{Facultad de Inform\'atica}
\department[1]{Instituto de Tecnolog\'ia del Conocimento (ITC)}
\institution{Complutense University of Madrid}
}
\email{mpmatias@ucm.es}

%

%
\begin{abstract}
Asymmetric multicore processors (AMPs) couple high-performance big cores and low-power small cores with the same instruction-set architecture but different features, such as clock frequency or microarchitecture.  Previous work has shown that asymmetric designs may deliver higher energy efficiency than symmetric multicores for diverse workloads. Despite their benefits, AMPs pose significant challenges to runtime systems of parallel programming models. While previous work has mainly explored how to efficiently execute task-based parallel applications on AMPs, via enhancements in the runtime system, improving the performance of unmodified data-parallel applications on these architectures is still a big challenge. In this work we analyze the particular case of loop-based OpenMP applications, which are widely used today in scientific and engineering domains, and constitute the dominant application type in many parallel benchmark suites used for performance evaluation on multicore systems. We observed that conventional loop-scheduling OpenMP approaches  are unable to efficiently cope with the load imbalance that naturally stems from the different performance delivered by big and small cores.  

To address this shortcoming, we propose \textit{Asymmetric Iteration Distribution} (AID), a set of novel loop-scheduling methods for AMPs that distribute iterations unevenly across worker threads to efficiently deal with performance asymmetry. We implemented AID in \textit{libgomp} --the GNU OpenMP runtime system--, and evaluated it on two different asymmetric multicore platforms. Our analysis reveals that the AID methods constitute effective replacements of the \texttt{static} and \texttt{dynamic} methods on AMPs, and are capable of improving performance over these conventional strategies by up to 56\% and 16.8\%, respectively. 
\end{abstract}

%
%


%

%

%
\maketitle

\section{Introduction}\label{sec:intro}

Delivering energy efficiency has become one of the major challenges for current and future processor designs~\cite{hennessypatterson19}. To address this issue, heterogeneous architectures -- where different core types are coupled on the same platform for diverse and specialized use -- have emerged in the last decade. Among the existing heterogeneous architectures, we can distinguish various categories according to the degree of diversity in core types~\cite{survey-amps,li10}. At one extreme of the design space we find those architectures coupling a number of general-purpose cores with accelerators~\cite{accelerators,Caulfield2016} (such as platforms combining CPUs with GPUs/FPGAs), or with special-purpose processing units~\cite{Jouppi2018,cellbre}. On these platforms the various processing units/cores do not expose a common ISA (Instruction Set Architecture). As a result, substantial programming effort is usually required to fully tap into their potential~\cite{hennessypatterson19,cellbre_prog,camps}. At the other extreme of the design spectrum, we can find asymmetric single-ISA multicore processors (AMPs)~\cite{single-isa-perf}, which integrate a mix of complex high-performance big cores and power-efficient small cores. Our work focuses on this type of heterogeneous architecture, where the  general-purpose nature of the cores coupled with their shared ISA allows the execution of unmodified applications.

AMP architectures have drawn the attention of major hardware players; the ARM big.LITTLE processor~\cite{arm-big-little,odroid} --widely extended in the mobile market segment -- or the upcoming Intel Lakefield SoC~\cite{lakefield} are clear examples of commercial AMP products. Another platform that also leverages different core types with a shared ISA, but in the high performance computing arena, is the Sunway TaihuLight supercomputer~\cite{moreto_jpdc,taihulight}, which led the Top500 list in 2016 and 2017. Despite the energy efficiency benefits of AMP designs~\cite{single-isa-perf}, effectively dealing with the different performance delivered by heterogeneous cores to the various applications, still constitutes a significant challenge to the different layers of the system software, ranging from the operating system~\cite{camps,li10} to the runtime system~\cite{criticalityBL,moreto_jpdc}. This complicates the widespread adoption of AMPs as general-purpose computing platforms~\cite{colabgp,camps}.

In this work we explore how to efficiently execute unmodified data-parallel loop-based OpenMP applications (potentially optimized for symmetric multicores) on AMPs. OpenMP~\cite{openmp} is a well-established standard for parallel programming on shared-memory architectures. A broad variety of scientific and engineering applications have been parallelized using the parallel for constructs of OpenMP~\cite{openmparch}. Many of these applications are also widely used today as a tool for performance evaluation on symmetric and asymmetric multicore systems~\cite{speccpu2017,cho_pact18,utility-based-ac,camps,moreto_jpdc}. Moreover, they constitute the dominant application type in many parallel benchmark suites, such as NAS~\cite{npbomp}, Rodinia~\cite{rodinia} or SPEC OMP~\cite{spec_omp}. Given the extended use of this kind of programs, guaranteeing a proper functioning of the runtime system for their efficient execution on AMPs is paramount to avoid drawing misleading conclusions from experimental analyses. We should highlight that the OpenMP specification~\cite{openmp} also provides extensions for the creation of task-based parallel applications. Despite the fact that many legacy parallel programs do not exploit task parallelism, there already exists a large body of work that pursues the efficient execution of task-based parallel applications on asymmetric multicores~\cite{criticalityBL,aawsr,butko_17,moreto_jpdc,maleabletasks19}.

The performance of data-parallel OpenMP applications on AMPs may be highly sensitive to the underlying mechanism used by the runtime system to distribute iterations in a parallel loop among worker threads (aka \textit{loop scheduling})~\cite{moreto_jpdc}. In this paper, we demonstrate that conventional loop-scheduling approaches in OpenMP inherently cause load-imbalance --as a result of the runtime system's unawareness of the performance asymmetry on the platform -- and may introduce substantial overheads in attempting to cope with the higher performance delivered by big cores relative to small ones. To address these issues we propose three new loop-scheduling methods that exploit \textit{Asymmetric Iteration Distribution} (AID). The proposed methods, referred to as \texttt{AID-static}, \texttt{AID-hybrid} and \texttt{AID-dynamic}, are meant as natural replacements for the existing \texttt{static} and \texttt{dynamic} loop-scheduling strategies on AMPs. Because our proposed approaches are transparently applied by the OpenMP runtime system without modifying applications, they enable performance portability across different asymmetric multicore platforms.

Apart from the design of the three AID methods, other important contributions of our work are as follows:
\begin{itemize}
\item We implemented the proposed loop-scheduling methods in \textit{libgomp}, the GNU OpenMP library (runtime system), which comes with GCC. To enable the activation of \textit{any} loop-scheduling approach (including ours) on a higher number of parallel loops in unmodified OpenMP applications we carried out a subtle modification in the GCC compiler. Applications have to be recompiled to benefit from this support. 
\item We carried out a comprehensive experimental evaluation on two asymmetric multicore platforms: a system featuring an ARM big.LITTLE processor, and an x86 emulated AMP platform including frequency-scaled ``small'' cores. 
\item We compared the effectiveness of the AID schemes with that of conventional OpenMP loop-scheduling methods. To this end we employed a wide range of parallel benchmarks from the NAS, Rodinia and PARSEC suites. Our analysis reveals that \texttt{AID-static} and \texttt{AID-hybrid} constitute effective replacements for the \texttt{static} method, as they are able to outperform \texttt{static} across the board and deliver up to a 56\% relative performance improvement. Moreover, our \texttt{AID-dynamic} method also makes a good alternative to \texttt{dynamic} loop scheduling on AMPs, and it is capable to improve the performance of some applications (by up to 16.8\% on the ARM big.LITTLE platform). More importantly, \texttt{AID-dynamic} allows to reduce runtime overheads, which might, otherwise, negate the benefits of dynamic iteration distribution, especially in scenarios where running iterations on big cores provides little performance benefit relative to using small cores.
\end{itemize}

The remainder of the paper is organized as follows. Section~\ref{sec:motivation} motivates key design considerations of our proposals. Section~\ref{sec:related} analyzes and discusses the existing related work. Section~\ref{sec:design} outlines the design and implementation of AID. Section~\ref{sec:experiments} covers the experimental evaluation, and Section~\ref{sec:conclusions} concludes the paper.

\section{Motivation}\label{sec:motivation}

The scalability of computationally intensive loop-based OpenMP programs on symmetric CMPs (Chip Multicore Processors) is mainly limited by two factors. The first one is the existence of purely sequential execution phases in the application~\cite{pa-cf}, such as the code in between parallel loops -- typically executed by the master thread -- or specific code snippets inside parallel regions enclosed in a \texttt{MASTER} or  a \texttt{CRITICAL} section, which can be defined via specific OpenMP constructs.  The second factor is the load imbalance that may arise when distributing and running the iterations of a parallel loop by the various worker threads. The degree of load imbalance in a loop is sensitive to the kind of processing performed by the loop (problem specific), the loop-scheduling method applied by the runtime system, and the effects of the memory hierarchy (e.g., cache and memory-bandwidth contention)~\cite{loopscheduling}.

The OpenMP specification~\cite{openmp} defines various loop-scheduling strategies enabling to cope with different parallel loops. The programmer may enforce which one to use by means of the optional \texttt{schedule} clause of the \texttt{omp for} construct. Specifically, when the \texttt{static} schedule is used (the default choice in many OpenMP implementations), the iterations available are evenly distributed among threads. Because the assignment of iterations to threads is efficiently performed at the beginning of the loop, this introduces virtually no 
overhead from the runtime system~\cite{loopscheduling}. The \texttt{static} schedule, however, leads to load imbalance in the event that loop iterations have uneven computational cost. The \texttt{dynamic} schedule is usually better suited to this scenario, where threads dynamically ``steal'' iterations from a shared pool with a certain configurable \textit{chunk} (i.e. number of iterations removed from the pool each time) as they complete the assigned work. Unfortunately, the overhead of assigning iterations dynamically can be substantial, and the non-predictive behavior of this approach tends to degrade data locality~\cite{openmp-dynamic}.

\begin{figure}[tbp]
\centering 
\includegraphics[width=0.35\textwidth]{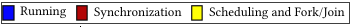}
\begin{subfigure}{0.28\textwidth}
	\includegraphics{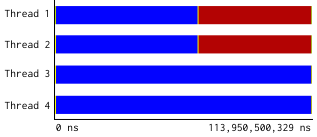}
	\caption{EP on 2B-2S}
	\label{fig:ep2b2s}
\end{subfigure}
\begin{subfigure}{0.28\textwidth}
	\includegraphics{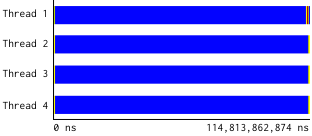}
	\caption{EP on 4S}
	\label{fig:ep4s}
\end{subfigure}
\vspace{-0.2cm}
\caption{Execution traces obtained with the Paraver tool~\cite{paraver} for the EP benchmark running with the static schedule and 4 threads with (a) two big and two small cores --2B-2S--, and (b) with four small cores --4S--.}
\end{figure}

The main goal of our work is to enable efficient execution of unmodified data-parallel OpenMP applications on asymmetric multicore systems. A critical challenge is how to address the load imbalance that naturally stems from to the different performance delivered by a big and a small core. To illustrate this fact, Fig.~\ref{fig:ep2b2s} shows an execution trace of the \texttt{EP} program --from NAS Parallel Benchmarks (NPB)--  running with 4 threads on an AMP configuration consisting of 2 big cores and 2 small cores. This program consists of a single parallel loop that spans the entire execution; iterations in this loop have roughly the same computational cost, so the \texttt{static} schedule (used in the experiment) is an acceptable choice on a conventional symmetric CMP. Due to the higher performance of big cores relative to small ones, big-core threads (1 and 2 in the figure) reach the implicit synchronization barrier sooner than small-core threads (3 and 4). This leads to poor utilization of big cores and low performance, as the completion of loop is bounded by the performance of small cores. Note that running the application on a CMP consisting of four small cores (Fig.~\ref{fig:ep4s}) delivers nearly the same performance than using two big cores and two small ones.

\begin{figure}[tbp]
\centering  
\begin{subfigure}{0.18\textwidth}
	\includegraphics[width=\textwidth]{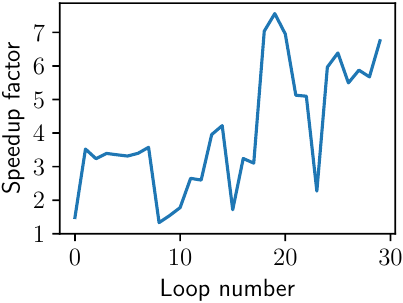}
	\caption{BT on Platform A}
	\label{fig:bta}
\end{subfigure}
\begin{subfigure}{0.18\textwidth}
	\includegraphics[width=\textwidth]{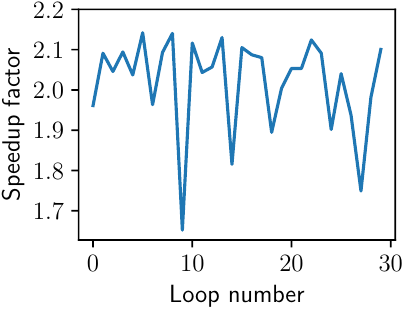}
	\caption{BT on Platform B}
	\label{fig:btb}
\end{subfigure}
\begin{subfigure}{0.18\textwidth}
 	\includegraphics[width=\textwidth]{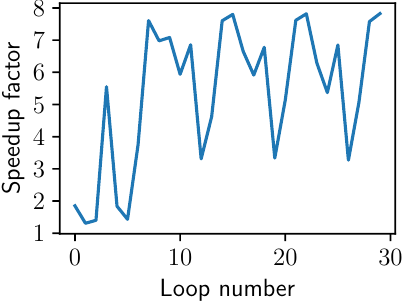}
	\caption{CG on Platform A}
	\label{fig:cga}
\end{subfigure}	
\begin{subfigure}{0.18\textwidth}
	\includegraphics[width=\textwidth]{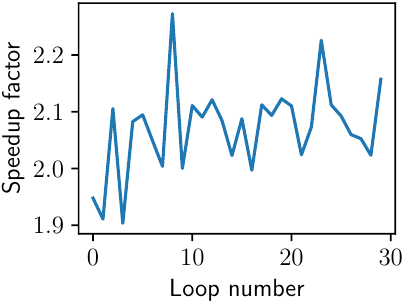}
	\caption{CG on Platform B}
	\label{fig:cgb}
\end{subfigure}
\vspace{-0.2cm}
\caption{Big-to-small relative performance for the first 30 loops of applications \texttt{BT} and \texttt{CG} on Platforms A and B.\label{fig:sfloops}}
\end{figure}

A potential way to mitigate the imbalance scenario highlighted in Fig.~\ref{fig:ep2b2s} would be to distribute iterations unevenly among threads on big and small cores, and in proportion to the relative performance across cores (aka \textit{speedup factor}~\cite{camps}). Note that the speedup factor is platform dependent, and largely depends on the nature of the code being executed~\cite{intel-amp,camps}. More importantly, we observed that the speedup factor may vary substantially across parallel loops of the same application. As an illustrative example, Fig.~\ref{fig:sfloops} shows the speedup factor for the first 30 loops of the \texttt{BT} and \texttt{CG} applications (from NPB) running on two AMP platforms used in our experiments, referred to as Platform A and Platform B respectively. More information on these platforms can be found in Sec.~\ref{sec:experiments}. To determine the speedup factor of each loop we ran the applications with a single thread on a big and on a small core and measured the completion time of individual loops. The figures report the ratio of these completion times for each loop. Clearly, the speedup factor varies greatly across loops in the same application, and  the trends observed on Platform A -- where loops run up to 7.7 times faster on a big core than on a small one-- differ substantially from those of Platform B. This suggests that using a single application-wide speedup factor to guide loop scheduling would not be enough to perform effective load balancing. In addition, informing the runtime system on the degree of performance asymmetry, by annotating each loop with the associated big-to-small speedup, would constitute an impractical approach towards achieving performance portability. In light of these observations, we advocate for dealing with performance asymmetry at runtime so as to deliver the benefits of AMPs to unmodified applications.

\section{Related Work}\label{sec:related}

Loop-scheduling is an active long-established research area, which has given rise to a plethora of proposals~\cite{loop_classical2,loop_classical1,donfack2012hybrid,hybrid_static_dyn,weighted_factoring,loopscheduling,kalesc2017,ompicpp19}.  Previous research has pointed out that a wide variety of loop-scheduling methods is necessary to meet the demands of an ample spectrum of applications~\cite{loopscheduling}. This observation has led researchers to propose additional scheduling methods~\cite{trapezoid,weighted_factoring,hybrid_static_dyn} to those already provided by the OpenMP standard. In addition, extensions in the specification of parallel loops have been considered to let the user define custom scheduling policies~\cite{ompicpp19}, which are especially beneficial to applications with irregular parallelism. 
Many of these proposals stand in contrast with our research, as we do not target specific application types but instead strive to provide efficient asymmetry-aware replacements for the well-known general-purpose \texttt{static} and \texttt{dynamic} loop-scheduling methods, which fail to cope with performance asymmetry. The implementation we created for the different variants of AID (asymmetric iteration distribution) on top of \textit{libgomp} enables the user to apply the AID methods without making changes to the application code. As we discuss in Sec.~\ref{sec:experiments} we also experimented with the \texttt{guided} method of OpenMP~\cite{openmp} but found that it delivers poor performance compared to \texttt{static} and \texttt{dynamic} on AMPs. For that reason, we focused on the creation of more efficient alternatives to the latter techniques. 

In the context of asymmetric multicores, previous experimental studies have been conducted to assess the effectiveness of the various loop scheduling methods in OpenMP~\cite{butko_16,moreto_jpdc}. Specifically, a very recent study~\cite{moreto_jpdc} demonstrates that \texttt{dynamic} is, in general, superior to \texttt{static} scheduling on AMPs, thanks to its ability to assign more work to threads running on big cores. Our experimental analysis, which considers a broader spectrum of loop-based OpenMP applications, allows us to arrive at a similar conclusion, but also reveals that the overhead of \texttt{dynamic} may negate the benefits of this loop-scheduling method in some cases. Moreover, unlike these previous works~\cite{butko_16,moreto_jpdc}, our main contribution is the design and implementation of new loop scheduling approaches for AMPs that outperform \texttt{static} and \texttt{dynamic}. Clearly, existing compiler-based solutions to deal with performance asymmetry~\cite{corr_18} are orthogonal to our AID methods, implemented entirely in the runtime system. 


Finally, it is worth noting that specific OS-level support~\cite{pa-cf} and hardware extensions~\cite{acs,utility-based-ac} have also been proposed to improve the performance of multithreaded applications on AMPs, by leveraging big cores as accelerators of different kinds of scalability bottlenecks present in parallel programs. Specifically, Saez et al.~\cite{pa-cf} propose a parallelism-aware scheduler that uses the runnable thread count (maintained by the OS) as a proxy for the amount of thread level parallelism in the application; this allows to detect program phases with limited parallelism and accelerate them using big cores. Notably, the same authors also show that  interaction between the runtime-system and the OS is paramount to reduce the possibility of wastefully running busy-waiting threads (e.g., waiting in a synchronization barrier) on big cores. As we discuss in Sec.~\ref{sec:multiapp}, our asymmetry-aware loop-scheduling policies could be integrated with asymmetry-aware OS schedulers like that one -- by means of explicit OS-runtime interaction -- to effectively adapt to multi-application scenarios, and thus benefit from automatic acceleration of serial execution phases~\cite{utility-based-ac,pa-cf}.

\section{Design and Implementation}\label{sec:design}

Our proposed loop-scheduling methods have been implemented in \textit{libgomp}, the GNU OpenMP runtime system. In particular we used the version of \textit{libgomp} that comes with GCC 8.3.0. 

The remainder of this section is organized as follows. We first explain the minor change we had to make to the GCC compiler to let the runtime system intervene in all parallel loops; this is crucial to activate our loop scheduling methods without modifying the application code. We then describe the motivation, design and implementation of the proposed approaches for AMPs. Finally, we discuss the potential interaction of the OS scheduler and the runtime system required to perform thread-to-core assignments and loop scheduling in a coordinated fashion in multi-application scenarios.  

\vspace{-0.2cm}
\subsection{Changes in the GCC compiler}\label{sec:gcc}

As stated earlier, the OpenMP \texttt{schedule} clause in the \texttt{omp for} construct is not mandatory. If a parallel loop's schedule clause is provided, API calls of the OpenMP runtime system are invoked by each worker thread when the loop begins (\texttt{GOMP\_loop\_<sched>\_start()}) and when it completes its alloted amount of iterations (\texttt{GOMP\_loop\-\_<sched>\_next()}), where \texttt{<sched>} denotes the loop-scheduling method selected. By contrast, when this clause is not provided in a loop, GCC employs the \texttt{static} schedule and removes loop-related runtime API calls in the generated code. Specifically, for those loops with no \texttt{schedule} clause, it  introduces code in the executable to conduct the static (even) iteration distribution without the interaction of the runtime system (deployed as a dynamic library). Note that the vast majority of loops in the OpenMP applications we used for the evaluation do not include the \texttt{schedule} clause. Most of these programs, when compiled with the \textit{vanilla} compiler, lack of loop-related runtime API calls. We can easily verify this by running the following command with the executable file of an OpenMP program (e.g., \texttt{bt.B}), whose loops omit the \texttt{schedule} clause:

{\fontsize{8.5}{8.5}\selectfont
\begin{verbatim}
$ nm -u bt.B | grep -i GOMP_
	U GOMP_barrier@@GOMP_1.0
	U GOMP_parallel@@GOMP_4.0
\end{verbatim}
}

As demonstrated in Sec.~\ref{sec:motivation} the \texttt{static} schedule may provide poor performance on AMPs, as it does not factor in performance asymmetry. When loop API calls are removed, even a modified version of the OpenMP runtime system (dynamic library) is unable to perform a different iteration distribution on an AMP; this is paramount to efficiently utilize the different core types. Because we aim to keep applications unmodified, we opted to make a simple change in the GCC compiler: we changed the default schedule mode for a parallel loop from \texttt{static} to \texttt{runtime}. As indicated in the OpenMP specification~\cite{openmp}, for those loops with this kind of schedule, the user can select the actual loop-scheduling method to be applied at runtime, by defining the \texttt{OMP\_SCHEDULE} environment variable with the desired value (e.g., \texttt{dynamic}). Note that this environment variable is read by the runtime system at the beginning of the execution, and the selected schedule is applied to all loops with the \texttt{runtime} schedule enabled. Compiling the aforementioned sample program with the modified GCC compiler does now include loop-related API calls in the associated executable file:   

{\fontsize{8.5}{8.5}\selectfont
\begin{verbatim}
$ nm -u bt.B_modified | grep -i GOMP_
	U GOMP_loop_end@@GOMP_1.0
	U GOMP_loop_end_nowait@@GOMP_1.0
	U GOMP_loop_runtime_next@@GOMP_1.0
	U GOMP_loop_runtime_start@@GOMP_1.0
	U GOMP_parallel@@GOMP_4.0
\end{verbatim}   
}  

We should highlight that compiling the applications in this way did not introduce noticeable overhead in the execution for the OpenMP programs we considered for the evaluation. In particular, we compiled each program with the same compiler switches (see Sec.~\ref{sec:experiments}), and using the original and modified version of the compiler. We observed that the execution of the programs built with the modified compiler and with \texttt{OMP\_SCHEDULE=static} do not exhibit apparent overheads relative to that associated with the binaries generated by the original compiler.  Notably, our subtle change in the compiler makes it now possible for the runtime system to intervene in the scheduling of \textit{all} parallel loops in the program.

\vspace{-0.3cm}
\subsection{Asymmetric Iteration Distribution (AID) Methods for AMPs}\label{sec:aid}

We implemented three new loop scheduling strategies for AMPs, which perform \textit{Asymmetric Iteration Distribution} (AID) among worker threads. We refer to these scheduling techniques as \texttt{AID-static},  \texttt{AID-hybrid} and \texttt{AID-dynamic}. The first two approaches are meant as a replacement of the \texttt{static} schedule on AMPs, and the third one is an alternative for the conventional \texttt{dynamic} schedule. Note that, in proposing these scheduling methods, we do not envision augmenting the OpenMP specification with new values for the \texttt{schedule} clause of a parallel for, but instead let the user activate and configure them on AMPs without making changes to the application code (e.g. via environment variables).

In designing the three scheduling techniques we leverage certain aspects of the existing lock-free implementation of the \texttt{dynamic} schedule in \textit{libgomp}. For the sake of clarity, we begin by describing how it is implemented. For each parallel loop, worker threads maintain a shared pool of iterations, whose state is represented via a set of fields in a \texttt{work\_share} shared structure. In particular, the \texttt{next} field (long integer) in that structure keeps track of the first iteration that has not been assigned to any thread yet; the \texttt{end} field holds the number of the last iteration in the loop. To steal \texttt{chunk}\footnote{In the \texttt{dynamic} schedule, \texttt{chunk} denotes the number of iterations that are repeatedly removed from the shared iteration pool by each thread. If the programmer does not specify the chunk value in the \texttt{omp for} construct, a default value of 1 is used. The runtime system maintains a scaled version of the chunk, whose value depends on the increment applied to the loop's control counter.} iterations from the shared pool (threads do so repeatedly until all iterations are completed), each thread invokes the \texttt{gomp\_iter\_dynamic\_next()} function. The actual stealing action is performed via the \texttt{fetch-and-add} instruction, used to atomically increment the \texttt{next} field by \texttt{chunk} units. The number of iterations assigned to the thread, which will be executed upon returning from the function, is determined by comparing the result of \texttt{fetch-and-add} and the \texttt{end} field of the \texttt{work\_share} structure.  

The new \textbf{AID-static} schedule is primarily meant for the efficient execution on AMPs of those parallel loops where all iterations require the same amount of work. Its goal is to distribute iterations so that threads assigned to big cores (henceforth, \textit{big-core threads}) and those mapped to small cores (henceforth, \textit{small-core threads}) complete their alloted iterations throughout the loop at nearly the same time, and by introducing as little overhead as possible from the runtime system (i.e. by reducing the number of runtime API calls). To this end the iterations are distributed unevenly, and in such a way that each thread is assigned a number of iterations proportional to the average relative performance observed when running a loop iteration on the core they actually run vs. that of a small core. To describe how to perform such a distribution, we first introduce the following notation:
\begin{itemize}
\item $NI$: total number of iterations in the parallel loop
\item $SF$: average speedup resulting from running an iteration of the loop on a big core relative to a small core. As shown in Sec.~\ref{sec:motivation} this value is loop specific. 
\item $N_B$ and $N_S$ is the number of application threads currently assigned to big and to small cores, respectively. We assume that (i) $N_B>0$ and $N_S>0$, (ii) no oversubscription exists (i.e. one thread per core on the platform at the most), and (iii) threads are not migrated between core types throughout the loop's execution.  
\item $k$ denotes the number of iterations ($0\leq{}k\leq{}NI$) of the loop allotted by \texttt{AID-static} to each small-core thread.
\end{itemize}

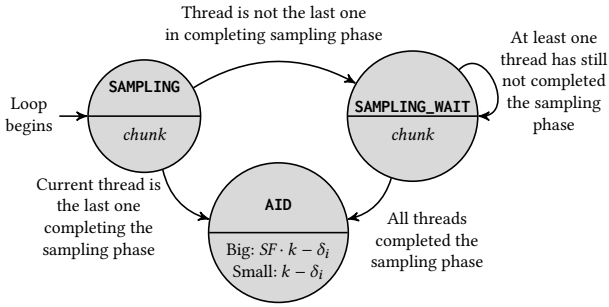
\begin{figure}[tbp]
\begin{tikzpicture}[scale=0.68,transform shape]

\def\minD{2cm}
\def\maxSepS{1.8*\minD}
\begin{scope}[->,>=stealth',auto,node distance=2.8cm,
                    semithick,scale=1.1,transform shape]
\tikzstyle{every state}=[text=black,circle split,fill=gray!30,draw=black,align=center,minimum width=\minD,inner sep=1pt]

   \node[state]  (sampling)  { {\bf\texttt{SAMPLING}} \\
    \nodepart{lower} \\ \textit{chunk}};    
   \node[state]  (wait) [right=\maxSepS of sampling.center]  {{\bf\texttt{SAMPLING\_} \\ {\bf\texttt{WAIT}}
    \nodepart{lower} \\  \textit{chunk}};    
   \node[state,inner sep=0]  (aid) at ($(sampling)!0.5!(wait) + (0,-0.8\minD)$) [below,inner sep=0]  { {\bf\texttt{AID}}  \\
    \nodepart{lower} \\ Big: $\mathrm{\textit{SF}}\cdot{}k - \delta_i$ \\ Small: $k - \delta_i$ };    
      
\path (sampling)  edge [bend left] node [yshift=0.2cm,align=center] {Thread is not the last one \\ in completing sampling phase} (wait)
(sampling)  edge [bend right] node [left,yshift=-0.3cm, xshift=-0.2cm,align=center] {Current thread is \\ the last one \\completing the \\ sampling phase} (aid)
		(wait)  edge [bend left] node[align=center, xshift=-0.2cm] {All threads \\ completed the \\  sampling phase} (aid)
		(wait)  edge [out=45,in=0,loop,looseness=2] node [right,align=center] {At least one \\ thread has still \\ not completed \\ the sampling \\ phase} (wait);

		;
		
\draw[<-,solid] (sampling.west) -- ++(-0.5cm,0)	 node [left, align=center] {Loop \\ begins};
\end{scope}
\end{tikzpicture}
\vspace{-0.2cm}
\caption{State diagram describing \texttt{AID-static}'s behavior. The lower part of each state node indicates the number of iterations that will be removed from the iteration pool by a thread in each case. \label{fig:aidstatic-diag}}
\end{figure}

\texttt{AID-static} assigns $SF\cdot{}k$ iterations to big-core threads. Because $\mathtext{NI}=N_{B}\cdot{}SF\cdot{}k+N_{S}\cdot{}k$, the following expression is used to determine $k$: $\frac{\mathtext{NI}}{N_{B}\cdot{}SF+N_{S}}$. Assuming that the runtime system does not have profiling information of the application, the $SF$ is unknown at the beginning of the loop. So, to perform the \texttt{AID-static} iteration distribution, the runtime must be equipped with a mechanism to determine the loop's $SF$ online. In our implementation, we introduce a \textit{sampling phase} at the beginning of the loop's execution, during which each worker thread runs \texttt{chunk} iterations. Once all threads have completed their sampling phase, we approximate the SF by dividing the average time that small-core threads take to complete the phase by the average completion time of the sampling phase in big-core threads. From this point on, the runtime system calculates $k$, and enforces that each small-core thread completes $k$ iterations, whereas big-core threads execute $SF\cdot{}k$ each. Note that to enforce the target iteration distribution throughout the loop, we keep track --for each thread $i$-- of the number of iterations it executed ($\delta{}_i$) before entering the $AID$ phase; $\delta{}_i$ is effectively subtracted from the corresponding final per-thread iteration assignment. Notably, this approach can be seamlessly extended to platforms with $NC$ core types ($NC\geq2$) as follows. For each group of threads on the same core type, a shared counter must be maintained to keep track of the average completion time of the sampling phase. Moreover, for each core type $j$, $SF_j$ must be measured, which denotes the relative performance (completion time ratio) of core $j$ relative to the one registered for the loop's chunk on the slowest core type ($j=1$). Let $N_j$ be the thread count on core type $j$; each thread in core type $j$ would receive $SF_j\cdot{}k$ iterations, where $k=\frac{NI}{\sum_{t=1}^{NC} N_t\cdot{}SF_t}$.

The state diagram in Fig.~\ref{fig:aidstatic-diag} depicts \texttt{AID-static}'s implementation, which we created by modifying the \texttt{gomp\_iter\_dynamic\_next()} function. Each edge in the graph represents an invocation to this function, which may trigger a transition between three possible thread states. Because \texttt{gomp\_iter\_dynamic\_next()} is invoked from each individual thread to remove iterations from the shared pool, state transitions occur independently for each thread. When the thread begins the loop execution, it automatically enters the \texttt{SAMPLING} state; in this state it completes \textit{chunk} iterations. The last thread in completing the sampling phase transitions automatically into the \texttt{AID} state; this thread is responsible for calculating $SF$ and $k$ --based on the completion times registered for each thread\footnote{To approximate a loop's $SF$ in a scalable fashion, we maintain two shared counters to keep track of the summation of execution times for sampling-phases in big-core and small-core threads, respectively. As soon as a thread completes the sampling phase it increments the associated counter atomically with the corresponding time. The average completion times for big and small cores --required to calculate the $SF$-- can be efficiently obtained by dividing the shared counters by $N_B$ and $N_S$, respectively.}-- and stores their values in the \texttt{work\_share} structure, shared among threads. The other threads transition instead into the \texttt{SAMPLING\_WAIT} state; during which they repeatedly steal \textit{chunk} iterations from the pool, until the last thread completes its sampling phase. At that point, they enter the AID state, where they receive the last iteration assignment (final invocation to \texttt{gomp\_iter\_dynamic\_next()}) based on the previous number of iterations completed ($\delta{}_i$), and on the core type they run on. We should highlight that the implementation of \texttt{AID-static} is lock free; we maintain an additional counter (updated atomically) in the \texttt{work\_share} structure to keep track of the number of threads that completed the sampling phase. Notably, the sampling phase --used in all AID methods-- has very low overhead (this is corroborated by our experiments in Sec 5C). Essentially two timestamps must be gathered for each thread to measure completion time; collecting timestamps is done very efficiently on Linux thanks to a vsyscall (the kernel shares memory with the process), so potentially costly syscalls are avoided. Note also that the simple calculations to determine \textit{SF} and $k$ are performed by one thread only, and while in the sampling phase, threads do complete iterations.

\begin{figure}[tbp]
\centering  
\includegraphics[width=0.35\textwidth]{legend_paraver.pdf}
\begin{subfigure}{0.22\textwidth}
	\includegraphics{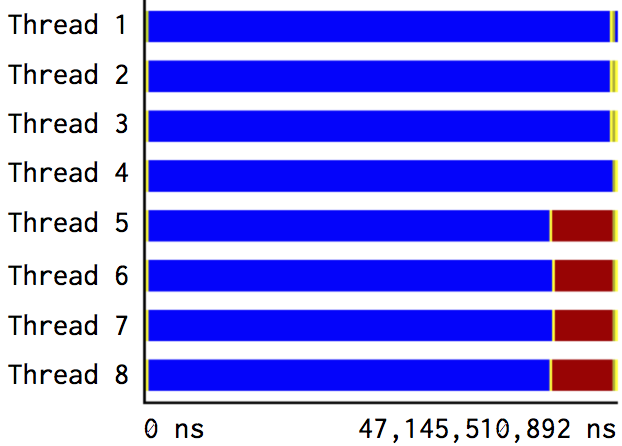}
	\caption{AID-static}
	\label{fig:epaid}
\end{subfigure}
\begin{subfigure}{0.22\textwidth}
	\includegraphics{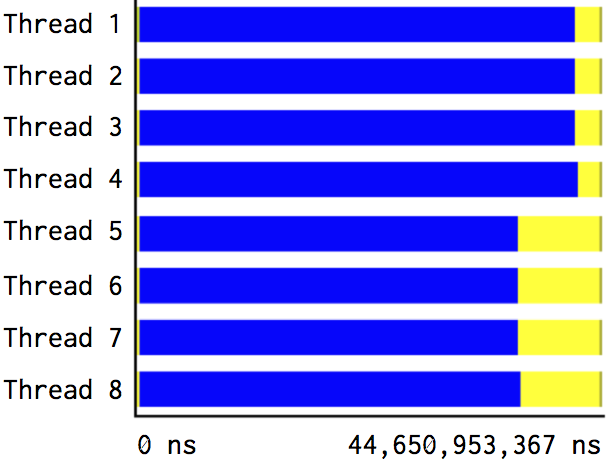}
	\caption{AID-hybrid (80\%)}
	\label{fig:epaidp}
\end{subfigure}
\begin{subfigure}{0.33\textwidth}
	\includegraphics{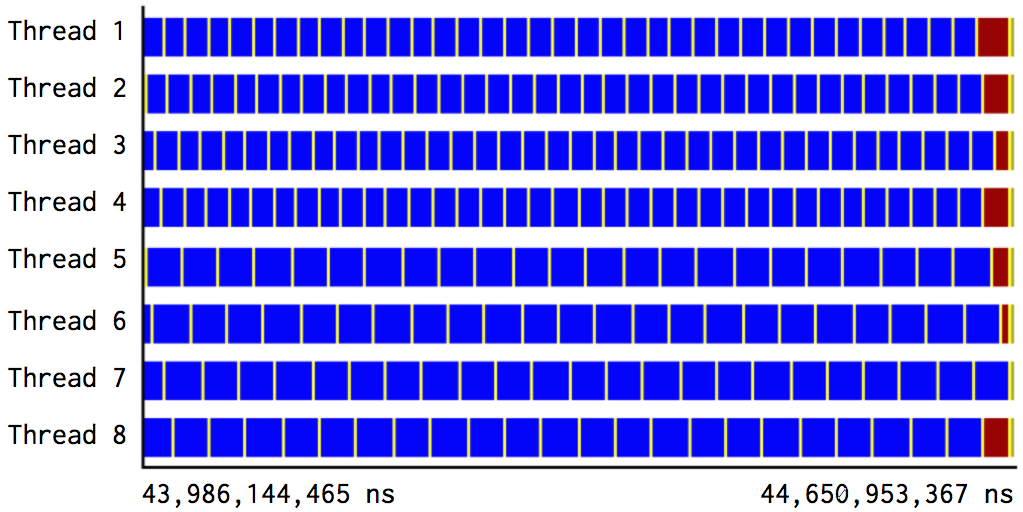}
	\caption{AID-hybrid (80\%) final part}
	\label{fig:epaidpd}
\end{subfigure}	
\vspace{-0.3cm}
\caption{Execution traces for the EP benchmark running with the \texttt{AID-static} and  \texttt{AID-hybrid} (80\%) schedules with 8 threads on Platform A.\label{fig:aid_ep}}
\end{figure}

We observed that the \texttt{AID-static} schedule provides satisfactory results for loops where all iterations require exactly the same amount of work. In the event that the workload differs slightly across iterations, \texttt{AID-static} may fail to balance the load. To illustrate this fact, Fig.~\ref{fig:epaid} shows an execution trace of the EP application on Platform A with \texttt{AID-static}, where the issue becomes apparent in its only parallel loop. In this case, the last four threads (5-8) are assigned to small cores, and complete their share of the parallel loop earlier than big-core threads. This occurs as a result of using the $SF$ approximated during the sampling phase, which in this case is not representative for the entire execution. 

To overcome this limitation, we propose \textbf{AID-hybrid}, a new scheme that combines \texttt{AID-static} with the conventional \texttt{dynamic} schedule of OpenMP. Specifically, \texttt{AID-hybrid} distributes only a fraction of the total loop iterations (percentage over $NI$) with \texttt{AID-static}; the remaining iterations are scheduled dynamically to try to balance the load at the end of the loop's execution, and at the expense of potentially higher runtime overhead. An execution trace for EP with \texttt{AID-hybrid} and percentage of 80\% (fraction of iterations distributed as in \textit{AID-static}) is shown in Fig.~\ref{fig:epaidp}. As it is evident, small-core threads (5 to 8) complete their share of the parallel loop earlier (as in Fig.~\ref{fig:epaid}), but in this case, they begin to remove iterations dynamically from the pool while big-core threads (1 to 4) complete their associated share. Despite the potential overhead introduced by the runtime system in the yellow region of the trace (i.e. where \texttt{dynamic} is engaged), \texttt{AID-hybrid} delivers a 10.5\% performance improvement over \texttt{AID-static}. This improvement stems from the lower degree of load imbalance delivered by \texttt{AID-hybrid}. The partial trace of Fig.~\ref{fig:epaidpd}, which captures the last 665 milliseconds of the execution, illustrates this fact.

The last loop-scheduling method we propose is \textbf{AID-dynamic}. With it we wanted to build an efficient replacement for the \texttt{dynamic} schedule specifically tailored to AMPs. To introduce less overhead than \texttt{dynamic}, \texttt{AID-dynamic} reduces the number of iteration-pool removals by using potentially larger chunks from big-core threads. It actually relies on two chunks defined by the user: \textit{Major} ($M$) and \textit{minor} ($m$), where $M\geq{}m$. \texttt{AID-dynamic} combines phases where all threads steal the same number of iterations ($m$) from the pool, with \textit{AID phases} where iterations are distributed unevenly. Specifically, during \textit{AID phases}, small-core threads receive $M$ iterations, whereas big-core threads are alloted $R\cdot{}M$ iterations. $R$ is the relative progress made by big-core threads over small-core ones throughout the loop, and is continuously updated by the runtime system.    

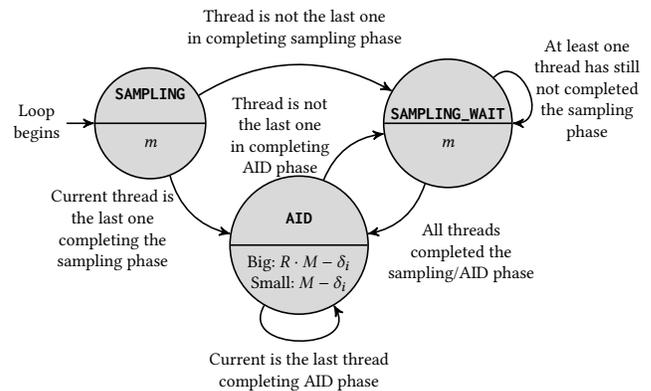
\begin{figure}[tbp]
\centering
\begin{tikzpicture}[scale=0.67,transform shape]

\def\minD{2cm}
\def\maxSepS{2.1*\minD}
\begin{scope}[->,>=stealth',auto,node distance=2.8cm,
                    semithick,scale=1.1,transform shape]
\tikzstyle{every state}=[text=black,circle split,fill=gray!30,draw=black,align=center,minimum width=\minD,inner sep=1pt]

   \node[state]  (sampling)  { {\bf\texttt{SAMPLING}} \\
    \nodepart{lower} \\ \textit{m}};    
   \node[state]  (wait) [right=\maxSepS of sampling.center]  {{\bf\texttt{SAMPLING\_} \\ {\bf\texttt{WAIT}}
    \nodepart{lower} \\  \textit{m}};    
   \node[state,inner sep=0]  (aid) at ($(sampling)!0.5!(wait) + (0,-0.94\minD)$) [below,inner sep=0]  { {\bf\texttt{AID}}  \\
    \nodepart{lower} \\ Big: $R\cdot{}M - \delta_i$ \\ Small: $M - \delta_i$ };    
      
\path (sampling)  edge [bend left] node [yshift=0.2cm,align=center] {Thread is not the last one \\ in completing sampling phase} (wait)

		(sampling)  edge [bend right] node [left,yshift=-0.3cm, xshift=-0.2cm,align=center] {Current thread is \\ the last one \\completing the \\ sampling phase} (aid)
		
		(wait)  edge [bend left] node[align=center, xshift=-0.4cm] {All threads \\ completed the \\  sampling/AID phase} (aid)

		(aid)  edge [bend left] node[left,align=center, xshift=-0.2cm,yshift=0.2cm] {Thread is not \\ the last one \\ in completing \\ AID phase} (wait)

		(aid)  edge [out=240,in=-60,loop,looseness=2] node [below,yshift=-0.1cm,align=center] {Current is the last  thread \\ completing AID phase} (aid)

		(wait)  edge [out=45,in=0,loop,looseness=2] node [right,align=center] {At least one \\ thread has still \\ not completed \\ the sampling \\ phase} (wait);

		;
		
\draw[<-,solid] (sampling.west) -- ++(-0.5cm,0)	 node [left, align=center] {Loop \\ begins};
\end{scope}
\end{tikzpicture}
\vspace{-0.35cm}
\caption{State diagram describing \texttt{AID-dynamic}'s behavior. Note that we performed an optimization that is not reflected in this diagram. The runtime system automatically switches to the \texttt{dynamic}($m$) schedule, as soon as it detects that the number of iterations remaining to execute is no greater than $M\cdot{}\left(N_B+N_S\right)$. This greatly improves load balancing at the end of the loop. \label{fig:aiddynamic-diag}}
\end{figure}

Fig.~\ref{fig:aiddynamic-diag} depicts the inner workings of \texttt{AID-dynamic}. This state diagram is a variant of the one in Fig.~\ref{fig:aidstatic-diag} that incorporates two new state transitions, and reflect different chunk values on the various states. As in \texttt{AID-static}, every thread goes through an initial sampling phase, and enters the \texttt{AID} state as soon as all threads complete their sampling phase. A key difference with \texttt{AID-static} lies in that the AID state here starts a new thread sampling phase, which spans the execution of the \textit{AID phase} with the number of iterations that have just been alloted. When all threads complete that \textit{AID phase} (longer than the initial sampling phase), a new AID assignment is performed. This process is repeated until no more iterations remain in the pool. The value of $R$ in the first AID phase is $SF$, which is calculated in the same way as in \texttt{AID-static}. The value of $R$ used in subsequent AID phases is $R'\cdot{}\mathtext{SM}$, where $R'$ is the value of R used in the previous AID phase and $\mathtext{SM}$ denotes an smoothing factor. $\mathtext{SM}$ is calculated by dividing the average time that small-core threads take to complete the previous AID phase by the average completion time of the same phase in big-core threads.

\subsection{Dealing with multi-application scenarios}\label{sec:multiapp}

In this work we focused on the scenario where a parallel application runs alone on the platform and it uses all the cores available.\footnote{To effectively deal with oversubscription scenarios, specific changes would be required in AID's current implementation, such as factoring in the load of each core in distributing iterations, and using CPU time instead of real time for SF estimation.} Under these circumstances, the runtime system typically binds threads to cores to avoid undesired migrations carried out by the operating system~\cite{loopscheduling,moreto_jpdc}. For simplicity in our implementation of the three variants of AID, the runtime system relies on the assumption that the $N_B$ threads with a lower thread ID in the runtime (TID $\in \lbrace0\;..\;N_{B}-1\rbrace$) are mapped to big cores, and the remaining ones (TID $\in \lbrace N_{B}\;..\;N_{S}-1\rbrace$) are assigned to small cores. Iterations are then allotted to the various threads according this mapping convention. The user must define the \texttt{GOMP\_AMP\_AFFINITY} environment variable so that the runtime system automatically binds thread-to-cores to enforce the aforementioned assignment.

Other authors have considered the execution of multiple parallel applications on CMPs for the efficient utilization of the various cores without oversubscription~\cite{pa-cf,utility-based-ac,cho_pact18,camps}. In this scenario,  the operating system, or an intermediate layer of the system software (underneath the per-application runtime system~\cite{cho_pact18}), is responsible for driving thread-to-core assignments. To enable coordinated scheduling decisions between an asymmetry-aware OS scheduler and our modified runtime system (without explicit CPU bindings), a set of minimal changes would be necessary in the OS. First, the OS scheduler should allow the runtime system to know how many threads of the application are mapped to big cores at all times. A shared memory region between the OS and the runtime system could be used to efficiently exchange information between both agents, thus removing the need of system calls. Second, in populating big cores, the OS scheduler should favor threads with lower TIDs in the same application, to comply with the aforementioned mapping convention in AID. Third, the runtime system would also greatly benefit from notifications from the OS when an application thread is migrated between cores of different types (e.g. by delivering a signal to the process). That would give the runtime system opportunities to readjust the distribution of iterations dynamically to improve load balance, possibly by combining our work-sharing version of AID, with work-stealing techniques~\cite{hybrid_static_dyn,ompicpp19}. Evaluating the effectiveness of these coordinated decisions between the runtime system and the OS is an interesting avenue for future work.

\section{Experimental evaluation}\label{sec:experiments}

In our evaluation, we compare the effectiveness of the proposed loop scheduling methods -- \texttt{AID-static}, \texttt{AID-hybrid} and \texttt{AID-dynamic} -- with that of the conventional \texttt{static} and \texttt{dynamic} methods of OpenMP. Note that we also experimented with the OpenMP \texttt{guided} loop-scheduling method, which assigns iterations dynamically with a decreasing chunk. Nevertheless, as in ~\cite{butko_16} we observed that \texttt{guided} is not superior than  \texttt{dynamic} on AMPs. Moreover, for the wide range of applications that we considered, \texttt{guided} increases completion time by 44\% and 65\% on average relative to \texttt{static} and \texttt{dynamic}, and never outperforms both of these two approaches for any program. For that reason, and to keep our analysis to the point, we do not discuss the per-application results of \texttt{guided}.

To assess the benefits of the various approaches on diverse architectures, we employed two AMP configurations, referred to as Platform A and Platform B. Platform A is the Odroid-XU4 board~\cite{odroid}, which integrates a 32-bit ARM big.LITTLE processor consisting of four \textit{big} cores (Cortex A15) and four \textit{small} (Cortex A7). Table~\ref{table:platform} shows more information about this system. Each core of the big.LITTLE processor features a private L1 cache and shares a last-level (L2) cache with the other cores of the same type. Platform B is an 8-core performance-asymmetric multicore system emulated with a server platform featuring an Intel Xeon E5-2620 v4 processor. On this platform all cores share a 20MB/20-way LLC. We ``create'' four \textit{slow} cores on this system by means of a joint reduction of the frequency and the duty cycle of the cores. Specifically, \textit{slow} cores operate at the lowest power P-state attainable on this platform (1.2 Ghz) and at the 87.5\% of the maximum duty cycle. The remaining (\textit{fast}) cores operate at 2.1GHz and at a full duty cycle. Downscaling both the frequency and the duty cycle on slow cores enables us to get bigger fast-to-slow performance ratios than what can be obtained by reducing the frequency only. Note that the turbo boost feature was disabled during the experiments. On both platforms we run GNU/Linux, using the 4.14.165 kernel version, with official support for both systems.

\begin{table}[tbp]
\centering
\caption{Features of the Odroid-XU4 board\label{table:platform}}
\vspace{-0.2cm}
{\footnotesize
\begin{tabular}{|c||c|c|}
\hline
Processor Model(s) & Cortex A15 &       Cortex A7 \\ \hline
Core count & 4 & 4 \\ \hline
Processor Frequency & 2.0GHz & 1.5 GHz \\ \hline
Pipeline & Out-of-order & In-order \\ \hline
Last-Level Cache (L2) & 2MB/16-way & 512KB/8-way \\ \hline \hline
Main Memory & \multicolumn{2}{c|}{2GB LPDDR3 @ 933MHz} \\ \hline
\end{tabular}
}
\vspace{-0.3cm}
\end{table}

\begin{figure*}
\def\figheight{3.1cm}
\centering
\begin{tikzpicture}[every node/.style={inner sep=0pt}]
\node(one) [scale=0.9,transform shape] {\includegraphics[height=\figheight]{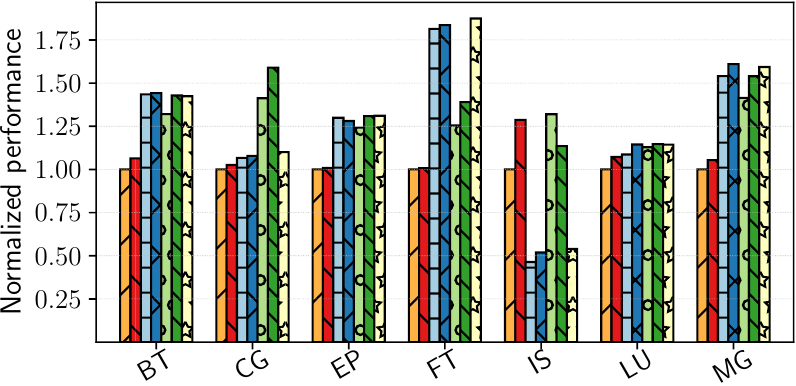}};
\node(two) [right=0cm of one.north east,below right] {\includegraphics[height=\figheight]{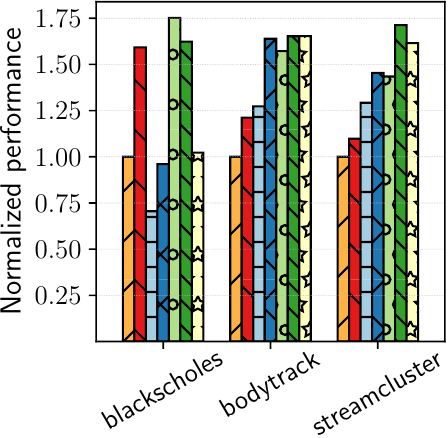}};
\node(three) [right=0cm of two.north east,below right] {\includegraphics[height=\figheight]{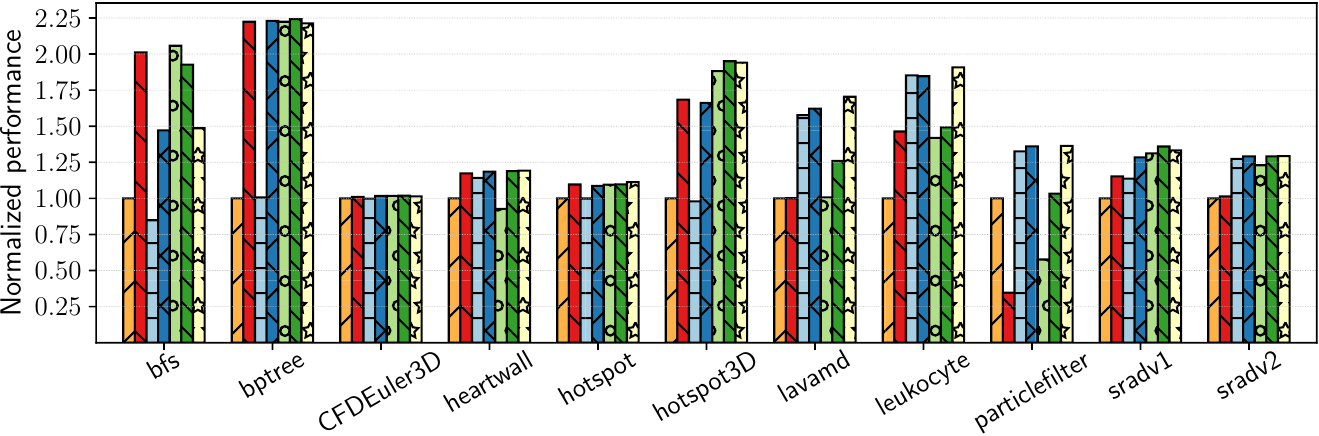}};
\node (legend) at ($(one.north west)!0.5!(three.north east)$) [above] {\includegraphics[width=0.8\textwidth]{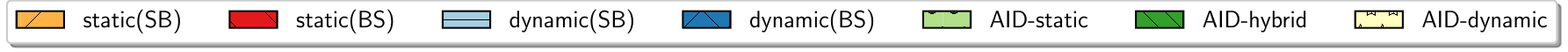}};
\end{tikzpicture}
\vspace{-0.65cm}
\caption{Normalized performance of the various applications obtained for the different OpenMP loop scheduling approaches on Platform A (Odroid-XU4 Board). Programs of different benchmark suites are displayed in separate subfigures.\label{fig:odroidres}}
\end{figure*}

For the experiments we employ 21 benchmarks from the NAS Parallel~\cite{npbomp}, PARSEC 3~\cite{parsec} and Rodinia~\cite{rodinia} benchmark suites. To activate the runtime system in all parallel loops, all programs were compiled with our modified version of GCC 8.3 using the \texttt{-O3} switch plus an additional option to apply platform-specific optimizations: \texttt{-mtune=cor\-tex-a15.cortex-a7} for Platform A, and \texttt{-mtune=broadwell} for Platform B. Since Rodinia applications have very short execution times with the default commands provided, we increased the input size of the benchmarks by following the approach proposed in~\cite{rodinia_commands}. For benchmarks from NAS parallel and PARSEC we used the B and the native input sets, respectively. Unfortunately, we could not gather results for a few OpenMP applications from PARSEC and NAS since they crashed on Platform A --in most cases due to exceeding the limited amount of physical memory available on that system (2GB). From those applications from Rodinia that built and ran successfully on that ARM platform, we report the results of 11 programs, where differences across the various loop-scheduling approaches were observed (others featured very low SF values or included very short parallel sections). In all experiments, we ran each program five times, and considered (for the experimental analysis) the geometric mean of the completion times of the application across the last four runs. The time of the first run of each program was discarded as it usually incurs extra overhead associated with bringing the program's input data into memory; the overhead is especially noticeable on Platform A, where the data is stored on an SD card.

\begin{figure*}
\def\figheight{3.1cm}
\centering
\begin{tikzpicture}[every node/.style={inner sep=0pt}]
\node(oned) [scale=0.9,transform shape] {\includegraphics[height=\figheight]{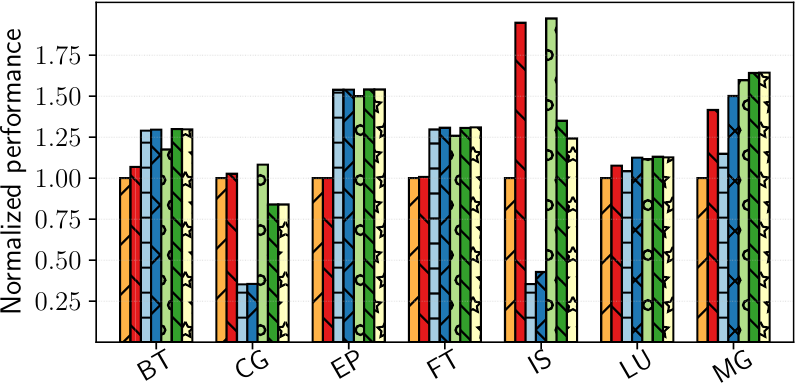}};
\node(twod) [right=0cm of oned.north east,below right] {\includegraphics[height=\figheight]{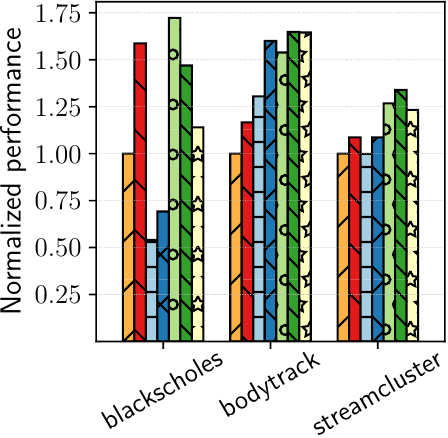}};
\node(threed) [right=0cm of twod.north east,below right] {\includegraphics[height=\figheight]{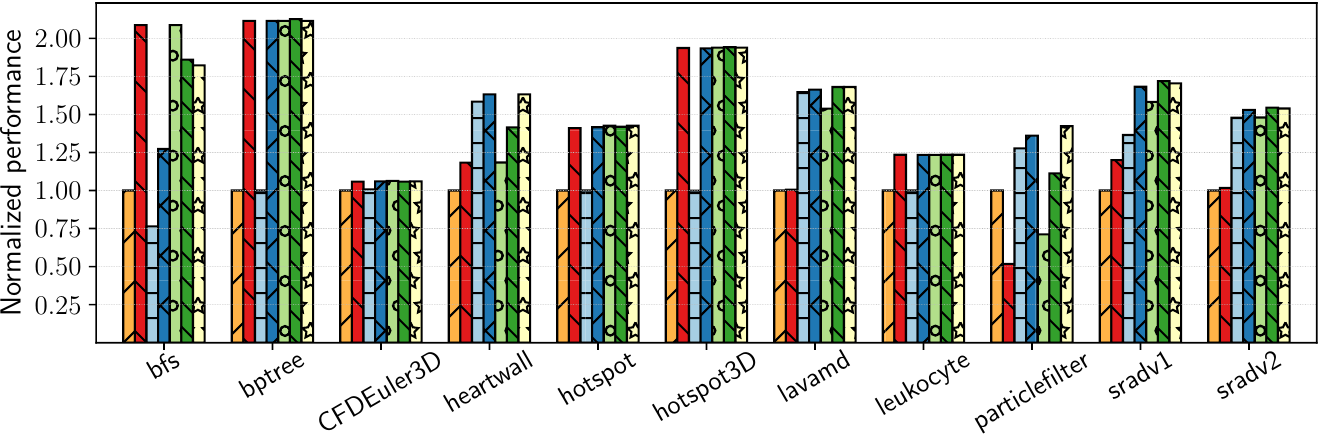}};
\node (legendd) at ($(oned.north west)!0.5!(threed.north east)$) [above] {\includegraphics[width=0.8\textwidth]{results_legend.pdf}};
\end{tikzpicture}
\vspace{-0.65cm}
\caption{Normalized performance of the various applications obtained for the different OpenMP loop scheduling approaches on Platform B (Xeon E5-2620 v4).\label{fig:intelres}}
\end{figure*}

\vspace{2pt}
\textit{A) Experiments with default chunk settings.} Figures ~\ref{fig:odroidres} and~\ref{fig:intelres} show the results obtained on Platforms A and B (respectively) for the various programs with the different loop scheduling techniques, and running with as many threads as the number of cores available (8). Table~\ref{table:mean} reports the average and geometric mean performance improvements of the different AID approaches relative to the conventional loop-scheduling approach they are meant to replace. Because the code of OpenMP applications was not modified, each loop scheduling technique explored is applied to all parallel loops in the program that do not include the \texttt{schedule} clause (this constitutes more than 95\% of the loops in the programs we used). On both platforms, big cores have CPU numbers ranging between 4 and 7; CPUs 0-3 are small cores. To isolate the performance benefit that comes from running serial execution phases --usually executed by the master thread (TID=0)-- on a big core, as well as other sources of imbalance for the \texttt{static} and \texttt{dynamic} method, we explored two approaches to bind threads to cores: $SB$ and $BS$. Under $SB$, CPU cores are populated in ascending order by thread ID, so threads 0-3 are assigned to small cores. By contrast, under $BS$, cores are assigned in descending order by thread ID, so big cores are reserved for threads 0-3. As explained in Sec.~\ref{sec:multiapp}, all variants of AID assume a $BS$ mapping in alloting iterations unevenly across threads, so $BS$ is always used in AID.

We begin by analyzing the results obtained on Platform A, which are displayed in Fig.~\ref{fig:odroidres}. The figures show the normalized performance of each approach, using \texttt{static(SB)} as the baseline. As in previous work~\cite{moreto_jpdc,butko_16}, we used here the default chunk values defined by the runtime system for \texttt{static} and \texttt{dynamic}. For the sampling phase we used a chunk of 1 for all AID approaches (referred to as the \textit{minor} ($m$) chunk in \texttt{AID-dynamic}). The performance numbers reported for \texttt{AID-hybrid} correspond to an execution with the same default chunk as \texttt{dynamic} (1), and where the uneven iteration distribution phase is applied to 80\% of the iterations. Later on we discuss the mechanism we used to determine that value, which constitutes a safe choice in our experimental platforms. In these experiments we used a \textit{Major} ($M$) chunk of 5 for \texttt{AID-dynamic}. Notably, in Sec. 5B we explore other chunk values for \texttt{dynamic} and \texttt{AID-dynamic}. 

By comparing the numbers associated with \texttt{static(BS)} and \texttt{static(SB)} we can see that substantial performance gains come in some cases from running the master thread on the big core. This is the case of \textit{IS}, \textit{blackscholes}, \textit{bfs} or \textit{hotspot3D}, with relative acceleration factors of up to 2.2x. In fact, in some programs, such as \textit{bptree}, the initialization phase (inherently sequential) takes the vast majority of the execution time; under these circumstances the observed acceleration comes primarily from running that serial phase (thread 0) on a big core. Conversely, when serial regions in the master thread constitute a low portion of the execution time, the \texttt{static} method clearly delivers lower performance than the remaining loop-scheduling approaches. This is caused by the performance asymmetry present on the platform, which introduces load imbalance, as we illustrated in Sec.~\ref{sec:motivation}.

Regarding conventional OpenMP scheduling approaches, the results underscore that a single method is not enough to cater to the demands of all applications. For example, for a few programs such as \textit{CG}, \textit{IS}, \textit{blackscholes} or \textit{bfs}, the \texttt{dynamic} method delivers poor performance, which is primarily caused by the high overhead it introduces. For these applications the overhead negates the benefits from assigning more work to big cores, delivering in some cases lower performance than the baseline. Of special attention is the case of \textit{IS} where the completion time increases by a factor of 1.93x relative to \texttt{static(SB)}. By contrast, for applications with parallel loops where iterations require an uneven amount of work, such as  \textit{FT}, \textit{leukocyte} or \textit{particlefilter}, the \texttt{dynamic} method is clearly beneficial. We would like to highlight the special case of \textit{particlefilter}, where \texttt{static(BS)} surprisingly delivers worse performance than \texttt{static(SB)}. We found that in this program, the final iterations in a long-running loop are more heavyweight computationally than the first iterations, so the \texttt{static} method under the $BS$ mapping actually assigns a higher amount of work to small cores. This is also the case for \texttt{AID-static}, hence the similar results. In this scenario, assigning iterations dynamically addresses the problem. Moreover, we observe that \texttt{dynamic} also deals in part with the load imbalance caused primarily from performance asymmetry across cores, such as in \textit{bodytrack}, \textit{sradv1} or \textit{sradv2}.

As for our proposed loop-scheduling methods, we can see that they generally constitute good replacements for the conventional approaches, thus we accomplish the main goal of our research. For simplicity in the discussion of the different variants of AID, we will focus exclusively on the comparison with the results of \texttt{static} and \texttt{dynamic} for the $BS$ mapping. As stated in Sec.~\ref{sec:design}, \texttt{AID-static} and \texttt{AID-hybrid} were designed to be natural alternatives on AMPs to the \texttt{static} method, meant for loops with equally-costly iterations. The results clearly reveal that \texttt{AID-static} outperforms \texttt{static} for the vast majority of workloads, delivering substantial relative performance gains for many benchmarks, such as for \textit{FT} (24.5\%), \textit{bodytrack} (29.7\%), or \textit{streamcluster} (30.7\%). Notably, the \texttt{AID-hybrid} approach is able to deliver even greater performance gains in these cases (up to 56\% for \textit{streamcluster}). These gains come from the uneven distribution of iterations, which effectively assigns more work to big-core threads than to small-core threads in proportion to the big-to-small relative speedup of the loop. 

\begin{table}[tbp]
\centering
\caption{Relative performance gains of the different AID variants\label{table:mean}}
\vspace{-0.2cm}
{\footnotesize
\begin{tabular}{c|c|c||c|c|}
\cmidrule{2-5} \\ [-0.36cm]
 & \multicolumn{2}{c||}{\textbf{Platform A}} & \multicolumn{2}{c|}{\textbf{Platform B}} \\ \hline
\textbf{Loop-scheduling schemes} & \textbf{Mean} & \textbf{Gmean} & \textbf{Mean} & \textbf{Gmean} \\ \hline
\texttt{AID-static} vs. \texttt{static(BS)} & 14.98\% & 13.54\% & 15.93\% & 14.64\% \\ \hline
\texttt{AID-hybrid} vs. \texttt{static(BS)} &  27.55 \% & 22.67 \% & 20.08\%  &16.06\% \\ \hline
\texttt{AID-dynamic} vs. \texttt{dynamic(BS)} & 3.12\% & 2.81\%  & 22.34\%   & 16.00\% \\ \hline
\end{tabular}
}
\vspace{-0.3cm}
\end{table}

For applications featuring parallel loops with an inherent source of load imbalance (i.e. also apparent in the execution on symmetric CMPs) assigning iterations dynamically is a better choice, so \texttt{AID-dynamic} provides higher performance in these cases. Note that applications that experience high relative performance gains with \texttt{dynamic} (e.g. \textit{FT}, \textit{leukocyte} or \textit{particlefilter}) also benefit from \texttt{AID-dynamic}. While for many programs both approaches perform in a 1\% range, \texttt{AID-dynamic} still provides noticeable performance gains w.r.t. \texttt{dynamic(BS)} for some applications, such as \textit{streamcluster} (11\%) and \textit{hotspot3D} (16.8\%). All in all, the results clearly demonstrate that \texttt{AID-dynamic} makes a good candidate for replacement of the conventional \texttt{dynamic} approach on AMPs, as it provides additional benefits that primarily come from making big-core threads and small-core threads steal a different number of iterations from the shared iteration pool based on the relative performance delivered by each core type for the loop in question.

We now turn our attention to the results of Platform B, shown in Fig.~\ref{fig:intelres}. In general, we observe very similar trends to those of Platform A (Fig.~\ref{fig:odroidres}), so the discussion provided thus far makes it possible to understand the reason of the main differences across loop-scheduling approaches. However, some additional observations are in order. First, on this platform, the big-to-small relative speedup of the various loops is different to that of Platform A (as shown in Sec.~\ref{sec:motivation}), so the normalized performance reported greatly differs in a few cases because of that. Second, the maximum big-to-small speedup observed across loops is substantially smaller on this platform (2.3x) than on Platform A (up to 8.9x). Because of the more reduced relative benefit from big cores, we expect that the overhead introduced by the runtime system --especially in dynamic loop-scheduling approaches-- may have a greater impact on performance, and potentially negate the benefits that come from dynamically assigning a higher share of iterations to big-core threads. The results clearly demonstrate that this is case for some applications such as \textit{CG}, \textit{IS} or \textit{blacksholes}, where the overhead introduced by the conventional \texttt{dynamic} approach is very high, leading to application slowdowns of up to 2.86x relative to the baseline (\textit{CG}). This underscores that the \texttt{dynamic} method could be potentially dangerous on those AMPs with a lower degree of performance asymmetry. Notably, \texttt{AID-dynamic} is able to reduce the runtime overhead (thanks to the AID phases), and it is still able to reap benefits for several applications, such as \textit{IS}, \textit{blacksholes} and  \textit{bfs}). Overall, on this platform \texttt{AID-dynamic} improves performance by 22\% on average relative to \texttt{dynamic}.           

\begin{figure*}
\centering
\includegraphics[width=1\textwidth]{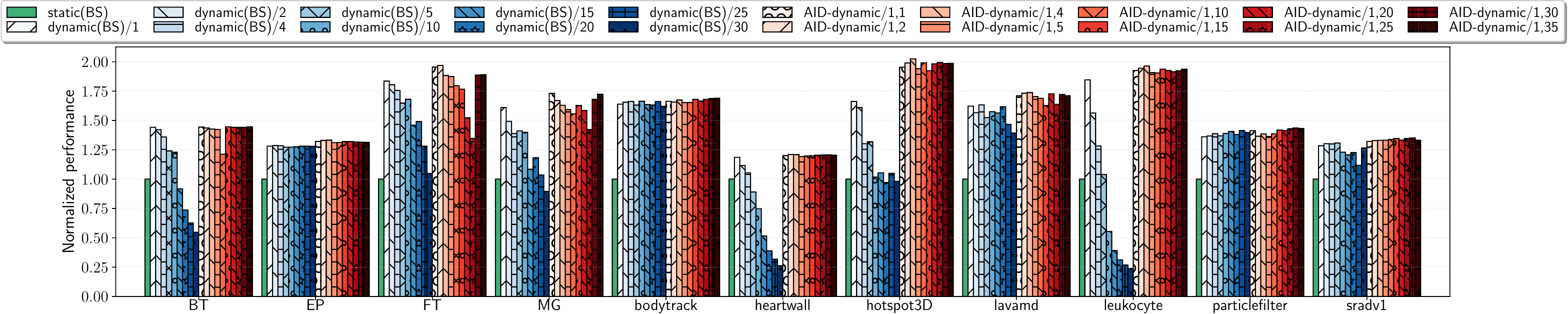}
\vspace{-0.5cm}
\caption{Normalized performance obtained with different chunk values for \texttt{dynamic} and various major (M) chunk values for \texttt{AID-dynamic} on Platform A. The number after ``/'' in the legend denotes the chunk setting for \texttt{dynamic} or the values of (m,M) --major and minor chunks-- for \texttt{AID-dynamic}. \label{fig:sweep}}
\end{figure*}

\begin{figure*}
\def\figheight{3.4cm}
\centering
\begin{tikzpicture}[every node/.style={inner sep=0pt},every label/.style={font=\bf\footnotesize}]
\node(oned) [scale=1,transform shape,label=below:{(a) Results Platform A}] {\includegraphics[height=\figheight]{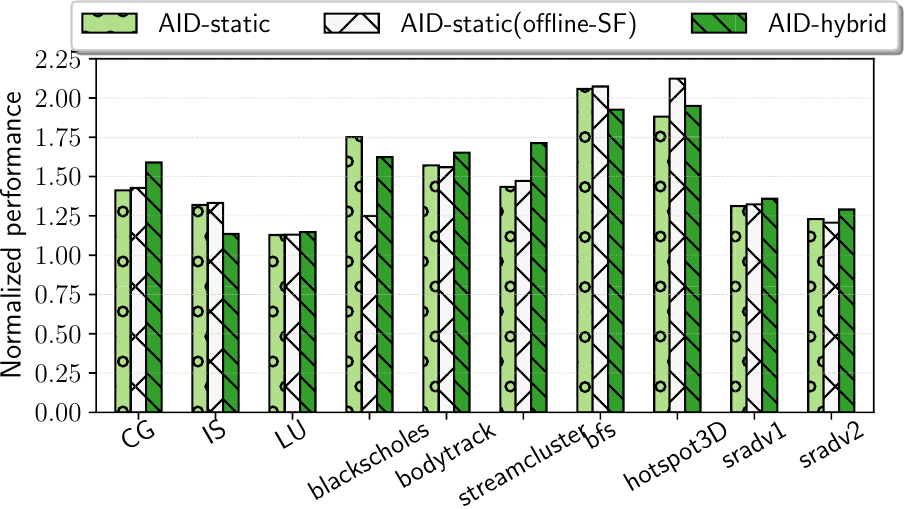}};
\node(twod) [right=0cm of oned.north east,below right,label=below:{(b) Results Platform B}] {\includegraphics[height=\figheight]{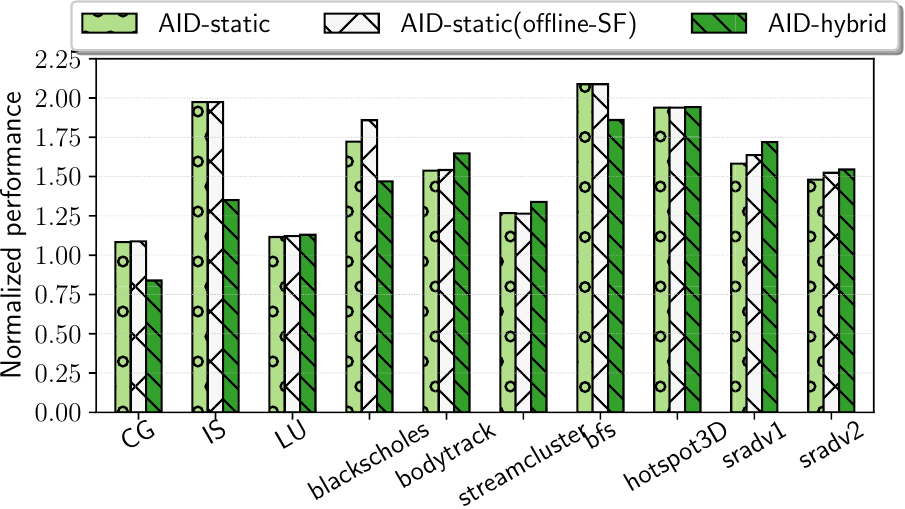}};
\node(threed) [right=0cm of twod.north east,below right,label=below:{(c) SF for \textit{blackscholes}  on Platform A}] {\includegraphics[height=3.3cm]{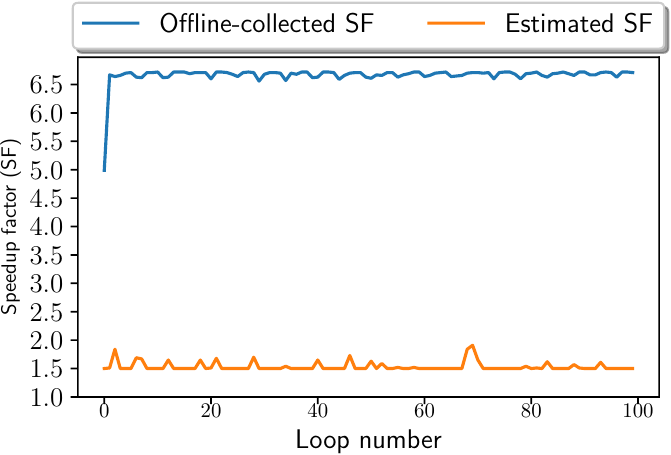}};
\end{tikzpicture}
\vspace{-0.25cm}
\caption{(a,b) Impact of SF prediction accuracy in performance, (c) Estimated vs. offline-gathered SF for \textit{blackscholes} on Platform A \label{fig:oracle}}
\end{figure*}

\vspace{2pt}
\textit{B) Experiments with different chunk/parameter values.} The performance of dynamic loop scheduling may be very sensitive to the choice of the chunk used~\cite{openmp-dynamic}. To assess how the chunk selection affects performance on AMPs we conducted additional experiments with applications that seem to benefit from distributing iterations dynamically (as observed in Fig.~\ref{fig:odroidres}). Fig.~\ref{fig:sweep} shows the results obtained for different settings of the chunk --for \texttt{dynamic}--  and the \textit{major} ($M$) chunk --for \texttt{AID-dynamic}. The results reveal that for many applications (e.g. \textit{BT}, \textit{FT} or \textit{leukocyte}) using bigger chunks clearly degrades performance. While using bigger chunks usually leads to lower runtime overhead, it may also cause load imbalance, especially at the end of a parallel loop; some threads may suddenly remove all remaining iterations from the pool leaving other threads with no work to do, thus degrading performance. While the choice of $M$ also has impact on \texttt{AID-dynamic}'s performance, the implementation of this loop-scheduling technique (see optimization described in Fig.~\ref{fig:aiddynamic-diag}) makes it possible to effectively remove this source of load imbalance, which makes this \texttt{AID-dynamic} less sensitive to the chunk choice. Overall, by comparing the performance  for \texttt{dynamic} and \texttt{AID-dynamic} with the best explored chunk setting for individual applications, we can observe that \texttt{AID-dynamic} improves dynamic by up to 21.9\% and by 5.5\% on average.

Note that we also conducted a sensitivity study to determine a suitable choice for the percentage used with \texttt{AID-hybrid}, a variant of \texttt{AID-static} where only that percentage of iterations is distributed among threads, and the remaining iterations are scheduled with \texttt{dynamic}. Due to space constraints, we are unable to display the corresponding figure with the results in the paper. Nevertheless, we summarize our main findings. The best choice for that percentage is application specific, and using a higher or lower percentage primarily depends on whether the application benefits from the \texttt{dynamic} method or not. Intuitively, applications like \textit{FT}, \textit{lavamd}, \textit{leukocyte} or \textit{particlefilter}, which are more favored by distributing iterations dynamically, benefit from lower percentages (60\% is a good choice for them). By contrast, parallel programs that experience substantial performance boost with \texttt{AID-static}, such as \textit{blackscholes}, benefit from high percentage values (90\% and above). All in all, we found that using 80\% with \texttt{AID-hybrid} provides a good trade-off between load balance and runtime overhead for the set of applications we used, so we opted to use it in our final experiments (values shown in Figs.~\ref{fig:odroidres} and~\ref{fig:intelres}).

\vspace{2pt}
\textit{C) Impact of potential SF-estimation inaccuracies.} The \texttt{AID-static} scheme relies on the ability to accurately estimate the average SF for each loop at runtime via a sampling phase. To assess the impact of potential SF-estimation inaccuracies in performance, we compared \texttt{AID-static} with a variant of that approach --referred to as \texttt{AID-static(offline-SF)}-- which is fed with the average SF values for each application loop, measured offline using the method described in Sec.~\ref{sec:motivation}. Under \texttt{AID-static(offline-SF)} the sampling phase is omitted; the runtime system just performs the asymmetric iteration distribution based on the loop's SF value provided. Figures~\ref{fig:oracle}(a) and~\ref{fig:oracle}(b) show the results obtained on Platforms A and B, respectively, for applications where \texttt{AID-static} or \texttt{AID-hybrid} provide comparable or better performance than \texttt{AID-dynamic}. The results reveal that \texttt{AID-static} is able to match the performance of \texttt{AID-static(offline-SF)} for many applications, and performs in a 3\% range of the latter for most programs. This suggests that the potential SF inaccuracies and the efficient implementation of the sampling phase have little impact on performance in most cases.

Of special attention is the case of \textit{blackscholes}. Whereas on Platform B \texttt{AID-static(offline-SF)} clearly outperforms \texttt{AID-static} (mainly to SF mispredictions), using offline-collected SF values leads to poor performance on Platform A. We found that this is due to  contention in the Last-Level Cache (LLC) --one per core type--, which may cause substantial performance degradation on current ARM big.LITTLE platforms~\cite{camps}. Because offline-collected SF values are gathered from single-threaded executions of the program, the performance degradation that may arise when multiple program threads compete for space in the LLC is not taken into consideration. In the case of \textit{blackscholes} the average per-thread LLC misses per 1K instructions registered for the 8-thread execution increases by a factor of 3.6x relative to the value of the single-threaded execution. This leads to severe performance degradation, and to low big-to-little performance ratios. In estimating the SF during the sampling phase, \texttt{AID-static} obtains lower (more realistic) SF values, which stand in contrast with the high SF values gathered offline (see Fig.~\ref{fig:oracle}(c)). Clearly, using the estimated values leads to better performance in this scenario. This underscores the importance of determining loop SF values at runtime.

\section{Conclusions and Future Work}\label{sec:conclusions}

In this paper we proposed AID, a set of three new loop-scheduling methods specifically tailored to data-parallel OpenMP applications on single-ISA asymmetric multicore processors (AMPs). All AID variants distribute iterations unevenly across worker threads to efficiently cater to performance asymmetry across cores. Two of the proposed schemes (\texttt{AID-static} and \texttt{AID-hybrid}) are meant as asymmetry-aware replacements for the conventional \texttt{static} OpenMP approach; the third approach --AID-dynamic-- was designed as an improved version of the existing \texttt{dynamic} loop scheduling method. We implemented our proposed strategies in \textit{libgomp} (the GNU OpenMP library), and introduced a seamless change in the GCC compiler to make it possible to activate the runtime system in all parallel loops without making changes to the application's code (applications just need to be recompiled).

Our evaluation, performed using two asymmetric multicore platforms (ARM and x86), shows that the \texttt{static} loop-scheduling strategy degrades performance as a result of the load imbalance that naturally arises due to the different benefit obtained from running code on a big core and on a small one. The \texttt{dynamic} method, by contrast, is able to deal with performance asymmetry more effectively as threads on big-cores can remove work faster from the shared pool (as they complete their previous assignment) than threads on small cores. Our experiments reveal that \texttt{AID-static} and \texttt{AID-hybrid} clearly outperform \texttt{static} by up to 30.7\% and 56\% respectively, and, in those cases where \texttt{dynamic} is capable of reaping benefits, \texttt{AID-dynamic} is able to match and even slightly improve the performance over \texttt{dynamic} (by up to 16.8\% on the ARM platform). Notably, we also found that the substantial runtime overheads introduced by \texttt{dynamic} in some cases may negate the benefits from completing more work on big cores; this is especially apparent in scenarios where the big-to-small speedup observed for the loop's code is modest (e.g. on the Intel platform). \texttt{AID-dynamic} is able to effectively reduce these overheads, making it possible to improve performance substantially over \texttt{dynamic}.

We should highlight that the reported performance benefits of our proposals come from applying the same loop-scheduling scheme to all loops in unmodified programs. We expect that further benefits can be obtained on AMPs by applying \texttt{AID-static} or \texttt{AID-hybrid} (replacements for \texttt{static}) to loops where iterations have the same amount of work, and \texttt{AID-dynamic} (replacement of \texttt{dynamic}) to the remaining loops. As this requires making changes in the application, we leave this analysis for future work, where we also plan to leverage existing approaches to help statically determine whether to use \texttt{static} or \texttt{dynamic} for specific loops~\cite{thoman2012} (that could help make a decision on what AID variant to use). Moreover, considering the case of multiapplication scenarios on AMPs, and designing variants of AID for other kind of heterogeneous architectures also constitute interesting avenues for future work.

\begin{acks}
This work was supported by the EU (FEDER), the Spanish MINECO and CM, under grants RTI2018-093684-B-I00 and S2018/TCS-4423. 
\end{acks}

%
\bibliographystyle{ACM-Reference-Format}
\bibliography{references}

\end{document}